\documentclass[twocolumn,showpacs,amsmath,amssymb]{revtex4}
\usepackage{graphicx}
\usepackage{dcolumn}
\usepackage{bm}
\usepackage{slashed}
\usepackage{amsmath}
\usepackage{float}
\usepackage{subfig}

\def\be{\begin{equation}}
\def\ee{\end{equation}}
\def\ba{\begin{array}}
\def\ea{\end{array}}
\def\bea{\begin{eqnarray}}
\def\eea{\end{eqnarray}}

\begin{document}
\title{Masses and Strong Decay properties of Radially Excited Bottom states B(2S)and B(2P) with their Strange Partners $B_{s}(2S)$ and $B_{s}(2P)$}

\author{\large  Pallavi Gupta}
\author{\large A. Upadhyay}
\affiliation{School of Physics and Materials Science, TIET, Patiala
- 147004, Punjab, INDIA}
\date{\today}
\baselineskip =1\baselineskip

\begin{abstract}

In this paper, we analyzed the experimentally available radially
excited charm mesons to predict the similar spectra for the n=2
bottom mesons. In the heavy quark effective theory, we explore the
flavor independent parameters $\Delta_{F}^{(b)} = \Delta_{F}^{(c)}$
and $\lambda_{F}^{(b)} = \lambda_{F}^{(c)}$ to calculate the masses
for the experimentally unknown n=2 bottom mesons B(2S), B(2P),
$B_{s}(2S)$ and $B_{s}(2P)$. We have also analyzed these bottom
masses by applying the QCD and $1/m_{Q}$ corrections to the
lagrangian leading to the modification of flavor symmetry parameters
as $\Delta_{F}^{(b)} = \Delta_{F}^{(c)}+\delta\Delta_{F}$ and
$\lambda_{F}^{(b)} = \lambda_{F}^{(c)}\delta\lambda_{F}$. Further
strong decay widths are determined using these calculated masses to
check the sensitivity of these corrections for these radially
excited mesons. The calculated decay widths are in the form of
strong coupling constant $\widetilde{g}_{HH}$, $\widetilde{g}_{SH}$
and $\widetilde{g}_{TH}$. We concluded that these corrections are
less sensitive for n=2 masses as compared to n=1 masses. Branching
ratios and branching fractions of these states are calculated to
have a deeper understanding of these states. These predicted values
can be confronted with the future experimental data.

\end{abstract}
\pacs{13.20.He, 12.39.Fe, 14.40.Nd } \keywords{Decays of bottom
mesons, Chiral Lagrangians, Bottom mesons $(|B|>0)$}
 \maketitle

\section{Introduction}
In the past years, the heavy-light mesons $(Q\overline{q})$ which
contain heavy quark Q and light anti-quark $\overline{q}$ have
received considerable experimental and theoretical attention due to
the existence of heavy quark and chiral symmetry. The study of heavy
light mesons provide a better understanding of non-perturbative
quantum chromodynamics (QCD). Many new discoveries have filled the
charm and bottom meson spectroscopy. The ground state charm mesons
such as $D(1S)$, $D(1P)$, $D_{s}(1S)$ and $D_{s}(1P)$ are well
established and are listed in particle data group \cite{pdg}. New
candidates for higher radial and orbital excitations in the charm
spectra include newly observed mesons like $D_{0}(2560)$,
$D^{*}_{1}(2680)$, $D_{2}(2740)$, $D^{*}_{3}(2760)$, $D_{J}(3000)$,
$D^{*}_{2}(3000)$ and $D^{*}_{J}(3000)$  and strange states
$D_{s1}(2860)$, $D_{s3}(2860)$ and $D_{s}(3040)$ \cite{lhcb2016,
lhcb2013, babar2010,5cs,6cs}. Non-strange charm states
$D^{*}_{J}(2460)$, $D_{J}(2560)$, $D^{*}_{J}(2680)$, $D_{J}(2740)$,
$D^{*}_{J}(2760)$, $D_{J}(3000)$ and $D^{*}_{J}(3000)$ are studied
in our previous work \cite{pallavi} where their $J^{P}$'s are
assigned as $1P_{\frac{3}{2}}2^{+}$, $2S_{\frac{1}{2}}0^{-}$,
$2S_{\frac{1}{2}}1^{-}$, $1D_{\frac{5}{2}}2^{-}$,
$1D_{\frac{5}{2}}3^{-}$, $2P_{\frac{1}{2}}1^{+}$ and
$2P_{\frac{1}{2}}0^{+}$ respectively. Observing the bottom
spectroscopy, it is realized that unlike the success in charm
sector, experimental information on higher excited bottom states is
scare. Till now, only ground state bottom mesons for $B(1S)$ and
excited P-wave mesons $B(1P)$ with $s_{l}$ = 3/2 along with their
strange partners are experimentally available \cite{249, 250, 8,
9,9a}. Recently in 2015, LHCb has observed new bottom mesons, which
have diverted theorists interest towards the bottom sector. LHCb
collaboration studied $B^{+}\pi^{-}$ and $B^{0}\pi^{-}$ mass
distributions by analyzing the p-p collisions at center-of-mass
energies of 7 and 8 TeV and observed four bottom states
$B_{1}(5721)$, $B_{2}^{*}(5747)$, $B_{J}(5840)^{0,+}$ and
$B_{J}(5960)^{0,+}$ \cite{10}. The results for $B_{J}(5960)^{0,+}$
are reported as:$M = 5969.20\pm 2.9\pm 5.1$ MeV and decay width
$\Gamma = 82.3\pm7.7\pm 9.4$ MeV Also in 2013, the CDF collaboration
studied the $B^{0}\pi^{+}$ and $B^{+}\pi^{-}$ mass distributions and
have observed neutral bottom state $B^{0}_{J}(5970)$ with mass $M =
5978 \pm 5 \pm 12$ MeV and decay width $\Gamma = 70^{+30}_{-20} \pm
30$ MeV \cite{11}. Since, this resonance decays in $B\pi$ final
state, this state is supposed to has a natural spin parity state.
The properties of $B_{J}(5970)$ state observed by CDF collaboration
are consistent with $B_{J}(5960)$ state observed by LHCb
collaboration, so they are assumed to be the same state.
Theoretically, these states are studied by using various models like
relativistic quark model\cite{12}, effective Lagrangian
approach\cite{13}, quark model\cite{14} etc. Being the fact that,
$B(5960)$ decays to $B\pi$ final states and its mass being close to
the mass of $2 ^{3}S_{1}$ state given in Ref's \cite{15, 16, 17},
this state is considered to belong to n=2, with $J^{P}$ = $1^{-}$
S-wave state in the bottom spectroscopy. $B_{1}(5721)$ and
$B_{2}^{*}(5747)$ are observed to belong to the bottom states $B(1
^{1}P_{1}(1^{+}))$ and $B(1 ^{3}P_{2}(2^{+}))$ respectively. Beside
these recently observed states, the information on the higher
orbital and radial excited
bottom states and their strange partners is still unknown. \\
So in this paper, we aim on predicting the properties of
experimentally missing radially excited bottom states $B(2S)$,
$B(2P)$ with their strange partners $B_{s}(2S)$ and $B_{s}(2P)$
states, which will be useful for both finding and understanding
these excited bottom mesons in future. We will enlighten some of the
properties like masses, strong decay widths, branching ratios,
branching fractions, strong coupling constants for these radial
excited bottom states. For this, we use Heavy Quark Effective Theory
(HQET)as our framework that includes heavy quark spin-flavor
symmetries making it invariant under $SU(2N_{f})$ transformations.
In the heavy quark limit for the heavy meson doublets, we expect
that the mass splittings among the different doublets and the
partial decay widths are independent of the heavy quark flavor
\cite{18}. The flavor symmetry implies that the spin averaged mass
splittings between the higher states and the ground state i.e.
$\Delta_{F}$ and the mass splittings between the spin partners of
the doublets i.e. $\lambda_{F}$ are flavor independent i.e.
$\Delta_{F}^{(b)} = \Delta_{F}^{(c)}$ and $\lambda_{F}^{(b)} =
\lambda_{F}^{(c)}$. We apply this heavy quark symmetry on the
experimental available data for n=2 charm mesons to predict the
properties of the corresponding bottom meson spectroscopy.
 We provide the mass spectra and strong decays for the bottom sector
  that will not only help in identifying the recent observed experimental
bottom mesons, but will also help in validating the authenticity of
the HQET model. This paper is organized as follows: Section 2 gives
the brief review of the HQET framework where we define the heavy
quark symmetry parameters and the possible QCD and $1/m_{Q}$
correction's to them. Section 3 represents the numerical analysis,
where we calculate the masses based on the heavy quark symmetry and
the corrections involved for the bottom states $B(1S)$, $B(1P)$,
$B_{s}(1S)$ and $B_{s}(1P)$. Next, we use these calculated masses as
an application, to predict the strong decay widths in terms of the
couplings. Section 4 presents the conclusion of our work.

\section{Heavy quark Effective Theory}

In heavy quark effective theory, spin and parity of the heavy quark
decouples from the light degrees of freedom as they interact through
the exchange of soft gluons only. Heavy mesons are classified in
doublets in relation to the total conserved angular momentum $i.e.$
$s_{l}=s_{\overline{q}}+l$, where $s_{\overline{q}}$  is the spin of
the light anti-quark and $\textit{l}$ is the orbital angular
momentum
 of the light degree of freedom. For $\textit{l} =0$ (S-wave) the
doublet is represented by $(P, P^{*})$ with $J^{P}_{s_{l}}=
(0^{-},1^{-})_{\frac{1}{2}}$, which for $\textit{l} =1$ (P-wave),
there are two doublets represented by $(P^{*}_{0},P^{'}_{1})$ and
$(P_{1},P^{*}_{2})$ with $J^{P}_{s_{l}}=(0^{+},1^{+})_{\frac{1}{2}}$
and $(1^{+},2^{+})_{\frac{3}{2}}$ respectively. These doublets are
described by the effective super-field $H_{a}, S_{a}, T_{a}$
\cite{18, 19, 20}, where the field $H_{a}$ describes the S-wave
doublets, $S_{a}$ and $T_{a}$ fields represents the P-wave doublets
for $J^{P}$ values $(0^{+},1^{+})_{\frac{1}{2}}$ and
$(1^{+},2^{+})_{\frac{3}{2}}$ respectively. Radially excited states
with radial quantum number n=2 are notated by $\widetilde{P},
\widetilde{P}^{*}$ and so on i.e. adding a $\sim$ symbol to the n=1
states. These fields are expressed as:
\begin{gather}
\label{eq:lagrangian}
 H_{a}=\frac{1+\slashed
v}{2}\{P^{*}_{a\mu}\gamma^{\mu}-P_{a}\gamma_{5}\}\\
S_{a}=\frac{1+\slashed
v}{2}\{P^{\mu}_{1a}\gamma_{\mu}\gamma_{5}-P^{*}_{0a}\}\\
T^{\mu}_{a}=\frac{1+\slashed v}{2}
\{P^{*\mu\nu}_{2a}\gamma_{\nu}-P_{1a\nu}\sqrt{\frac{3}{2}}\gamma_{5}
[g^{\mu\nu}-\frac{\gamma^{\nu}(\gamma^{\mu}-\upsilon^{\mu})}{3}]\}
\end{gather}

 HQET is developed by expanding
the QCD lagrangian in power of $1/m_{Q}$, in which heavy quark
symmetry breaking terms are studied order by order.  Applying finite
heavy quark mass corrections, HQET lagrangian to order of $1/m_{Q}$
is
\begin{gather}
\label{eq:lagrangian}
\mathcal{L}=\overline{h}_{v}(iv.D)h_{v}+\overline{h}_{v}\frac{(iD_{\bot})^{2}}{2m_{Q}}h_{v}+\overline{h}_{v}\frac{g\sigma_{\mu\nu}G^{\mu\nu}}{4m_{Q}}h_{v}+\textrm{O}(\frac{1}{m^{2}_{Q}})
\end{gather}
Where, $D^{\mu}_{\bot}=D^{\mu}-v^{\mu} v.D$ is orthogonal to heavy
quark velocity v, and
$G^{\mu\nu}=T_{a}G_{a}^{\mu\nu}=\frac{\imath}{g_{s}}[D^{\mu},D^{\nu}]$
is the gluon field strength tensor. In the limit
$m_{Q}\rightarrow\infty$, only first term $\overline{h}(iv.D)h$
survives. The second term $D^{2}_{\bot}$ is arising from the off
shell residual momentum of the heavy quark in the non relativistic
model and it represents the heavy quark kinetic energy
$\frac{p^{2}_{Q}}{2m_{Q}}$\cite{20a}. This term breaks the flavor
symmetry because of the explicit dependence on $m_{Q}$, but does not
break the spin symmetry of the HQET. The third term in the above
equation i.e. $g\sigma_{\mu\nu}G^{\mu\nu}$ represents the magnetic
moment interaction coupling of the heavy quark spin to the gluon
field. This term breaks both the flavor and spin symmetry and is
known as chromo-magnetic term.  The mass of the heavy-light hadron
to the first order of $1/m_{Q}$ is represented as :
\begin{gather}
\label{eq:lagrangian}
M_{X}=m_{Q}+\overline{\Lambda}-\frac{\lambda_{1}}{2m_{Q}}-d_{H}\frac{\lambda_{2}}{2m_{Q}}
\end{gather}
In this equation, $d_{H} =-4(S_{Q}.S_{l})$ is the Clebsch factor,
with $d_{H}$ = -3 or 1 for J = 0 or 1 respectively for $S_{q}$ = 1/2
and $d_{H}$ = -5 or 3 for J =  1 or 2 for $S_{q}$ = 3/2. The
$\lambda_{1}, \overline{\Lambda}$ and $\lambda_{2}$ are the
non-perturbative parameters, whose values are of the order of
$\Lambda^{2}_{QCD}$ where,
\begin{gather}
\overline{\Lambda} \equiv  \frac{1}{2}\langle H^{(Q)}\mid H_{0} \mid
H^{(Q)}\rangle
\end{gather}

\begin{gather}
\label{eq:lagrangian}
 \lambda_{1} = \frac{\langle
H_{Q}|\overline{Q}(iD_{\bot})^{2}Q|H_{Q}\rangle} {2m_{H_{Q}}}
\end{gather}

\begin{gather}
\label{eq:lagrangian}
 \lambda_{2} = \frac{\langle
H_{Q}|\overline{Q}\frac{1}{2}\sigma.G
Q|H_{Q}\rangle}{2d_{H}m_{H_{Q}}}
\end{gather}

The parameter $\overline{\Lambda}$ is the energy of the light quark
fields (i.e. brown muck), $\lambda_{1}$ term represents the kinetic
energy of the heavy quark Q and the term $\lambda_{2}$ gives the
chromomagnetic interaction energy \cite{21,22,23}. Since value of
kinetic energy of the heavy quark is positive , the value of the
parameter $\lambda_{1}$ should be negative. $\overline{\Lambda}$ is
the HQET parameter whose value is same for all the particles in a
spin-flavor multiplet. $\overline{\Lambda}$ does not depend on the
light quark flavor if there is $SU(3)$ symmetry, but for the
breaking of this symmetry $\overline{\Lambda}$ is different for
strange and non-strange heavy -light mesons and is denoted by
$\overline{\Lambda}_{s}$ and $\overline{\Lambda}_{u,d}$
respectively. \\In limit $m_{Q}\rightarrow\infty$, only the first
term of the HQET lagrangian (equation 4) will have the effect of
interaction. Based on various fields defined in equations 1-3, the
kinetic terms of the heavy meson doublets and of the $\Sigma$ field
of light pseudo-scalar mesons, present in the effective Lagrangian
\cite{18} are as:
\begin{multline}
\mathcal{L} = i Tr[\overline{H_{b}}v^{\mu}D_{\mu
ba}H_{a}]+\frac{f_{\pi}^{2}}{8}Tr[\partial^{\mu}\Sigma\partial_{\mu}\Sigma^{\dag}]\\
+Tr[\overline{S_{b}}(iv^{\mu}D_{\mu ba}-\delta_{ba}\Delta_{S})S_{a}]\\
+Tr[\overline{T_{b}}^{\alpha}(iv^{\mu}D_{\mu
ba}-\delta_{ba}\Delta_{T})T_{\alpha a}]
\end{multline}
where the operator D is given as:
\begin{equation}
\begin{split}
D = -\delta_{ba}\partial^{\mu}+ \mathcal{V}_{\mu ba}\\
= -\delta_{ba}\partial^{\mu}+
\frac{1}{2}(\xi^{\dag}\partial_{\mu}\xi+\xi\partial_{\mu}\xi^{\dag})ba
\end{split}
\end{equation}

 The mass parameter
$\Delta_{F}$ describes the mass splitting between the higher mass
doublets (F) and the ground state H field doublet. This mass
parameter $\Delta_{F}$ (F = S,T) can be written in terms of the spin
average mass of these doublets as:
\begin{gather}
\label{eq:lagrangian} \Delta_{F} =
\overline{M}_{F}-\overline{M}_{H}, F = S,T
\end{gather}
where
\begin{gather}
 \overline{M}_{H}=(3m^{(Q)}_{H^{*}}+m^{(Q)}_{H})/4\\
\overline{M}_{S}=(3m^{(Q)}_{S^{*}}+m^{(Q)}_{S})/4 \\
\overline{M}_{T}=(5m^{(Q)}_{T^{*}}+3m^{(Q)}_{T})/8
\end{gather}

The mass degeneracy between the members of the doublets breaks at
the $1/m_{Q}$ corrections to the heavy quark limit. This correction
is of the form of
\begin{equation}
\begin{split}
\mathcal{L}_{1/m_{Q}} =
\frac{1}{2m_{Q}}{}\lambda_{H}Tr[\overline{H}_{a}\sigma^{\mu\nu}H_{a}\sigma_{\mu\nu}]\\
-\lambda_{S}Tr[\overline{S}_{a}\sigma^{\mu\nu}S_{a}\sigma_{\mu\nu}]\\
+\lambda_{T}Tr[\overline{T}^{\alpha}_{a}\sigma^{\mu\nu}T_{a}^{\alpha}\sigma_{\mu\nu}]
\end{split}
\end{equation}
where $\lambda_{H}, \lambda_{S}$ and $\lambda_{T}$ are the hyperfine
splittings between the spin partners in each doublet with

\begin{gather} \label{eq:lagrangian}
\qquad \qquad \qquad \lambda_{H} =\frac{1}{8}(M_{P^{*}}^{2}-M_{P}^{2})\\
\lambda_{S} = \frac{1}{8}(M_{P^{'}_{1}}^{2}-M_{P_{0}^{*}}^{2})\\
\lambda_{T} = \frac{3}{16}(M_{P^{*}_{2}}^{2}-M_{P_{1}}^{2})\qquad
\qquad \qquad
\end{gather}

Flavor symmetry implies
\begin{gather}
\label{eq:lagrangian} \Delta_{F}^{(c)} = \Delta_{F}^{(b)}\\
\lambda_{F}^{(c)} = \lambda_{F}^{(b)}
\end{gather}

i.e. $\Delta_{F}$ parameter which represents the mass splittings
between the higher mass doublets and the ground state doublet and
the $\lambda_{F}$ parameter which  represents the hyperfine
splittings between spin partners, are same for both charm and bottom
mesons.

This symmetry is broken by the higher order terms in the HQET
lagrangian involving terms of factor $1/m_{Q}$ and the parameters
$\Delta_{F}$ and $\lambda_{F}$  are modified by extra term
$\delta\Delta_{F}$ and $\delta\lambda_{F}$ \cite{21}.

The hyperfine splitting term $\lambda_{F}$ which originates from the
chromomagnetic interaction is dominated by the QCD corrections and
the $1/m_{Q}$ effect is neglected\cite{22}. QCD corrections changes
the $\lambda_{F}$ relation to
\begin{gather}
\label{eq:lagrangian} \lambda_{F}^{(b)} =
\lambda_{F}^{(c)}(\frac{\alpha_{s}(m_{b})}{\alpha_{s}(m_{c})})^{9/25}
\end{gather}

Difference of the spin averaged masses at $1/m_{Q}$ order is
\begin{gather}
\overline{M}_{S}-\overline{M}_{H} =
\overline{\Lambda}_{S}-\overline{\Lambda}_{H}-\frac{\Lambda_{1}^{S}}{2m_{Q}}+\frac{\Lambda_{1}^{H}}{2m_{Q}}
\end{gather}
which modifies the parameter $\Delta_{F}$ by $\delta \Delta_{F}$,
i.e.
\begin{equation}
\begin{split}
\Delta_{F}^{(b)} = \Delta_{F}^{(c)}+ \delta \Delta_{F},\\
 \text{with}\qquad \qquad
\delta \Delta_{F}
  = (\Lambda_{1}^{F} - \Lambda_{1}^{H})[\frac{1}{2m_{c}}-\frac{1}{2m_{b}}]
\end{split}
\end{equation}
\\ Also in the heavy quark effective theory, the strange states possess the property, that the effect of the
strange quark is to shift the mass of a given state by the same
amount in the fundamental mode n=1 and for n = 2 radial excitation
doublet.
\begin{gather}
\label{eq:lagrangian}
 M_{P_{S}} - M_{P} = \widetilde{M}_{P_{S}} - \widetilde{M}_{P}
\end{gather}
 Masses of the heavy hadrons can be used to calculate other properties
like strong decays, radiative decays, magnetic moments etc. Strong
interactions are very important for the study of heavy hadrons
containing one heavy and one light quark in the non-perturbative
regime.

At the leading order approximation, using the lagrangians $L_{HH},
L_{SH}, L_{TH}, L_{YH}, L_{ZH}$, the two body strong decays of
$Q\overline{q}$ heavy-light charm mesons are given as
\\$(0^{-},1^{-}) \rightarrow (0^{-},1^{-}) + M$
\begin{gather}
\label{eq:lagrangian} \Gamma(1^{-} \rightarrow 1^{-})=
C_{M}\frac{g_{HH}^{2}M_{f}p_{M}^{3}}{3\pi f_{\pi}^{2}M_{i}}\\
\Gamma(1^{-} \rightarrow 0^{-})=
C_{M}\frac{g_{HH}^{2}M_{f}p_{M}^{3}}{6\pi f_{\pi}^{2}M_{i}}\\
\Gamma(0^{-} \rightarrow 1^{-})=
C_{M}\frac{g_{HH}^{2}M_{f}p_{M}^{3}}{2\pi f_{\pi}^{2}M_{i}}
\end{gather}

 $(0^{+},1^{+}) \rightarrow (0^{-},1^{-}) + M$
\begin{gather}
\label{eq:lagrangian} \Gamma(1^{+} \rightarrow 1^{-})=
C_{M}\frac{g_{SH}^{2}M_{f}(p^{2}_{M}+m^{2}_{M})p_{M}}{2\pi f_{\pi}^{2}M_{i}}\\
\Gamma(0^{+} \rightarrow 0^{-})=
C_{M}\frac{g_{SH}^{2}M_{f}(p^{2}_{M}+m^{2}_{M})p_{M}}{2\pi
f_{\pi}^{2}M_{i}}
\end{gather}

 $(1^{+},2^{+}) \rightarrow (0^{-},1^{-}) + M$
\begin{gather}
\label{eq:lagrangian} \Gamma(2^{+} \rightarrow 1^{-})=
C_{M}\frac{2g_{TH}^{2}M_{f}p_{M}^{5}}{5\pi f_{\pi}^{2}\Lambda^{2}M_{i}}\\
\Gamma(2^{+} \rightarrow 0^{-})=
C_{M}\frac{4g_{TH}^{2}M_{f}p_{M}^{5}}{15\pi f_{\pi}^{2}\Lambda^{2}M_{i}}\\
\Gamma(1^{+} \rightarrow 1^{-})=
C_{M}\frac{2g_{TH}^{2}M_{f}p_{M}^{5}}{3\pi
f_{\pi}^{2}\Lambda^{2}M_{i}}
\end{gather}
In the above decay widths, $M_{i}$ and $M_{f}$ stands for initial
and final meson mass, $p_{M}$ and $m_{M}$ are the final momentum and
mass of the light pseudo-scalar meson respectively. The coefficient
$C_{\pi^{\pm}}, C_{K^{\pm}}, C_{K^{0}}, C_{\overline{K}^{0}}=1$,
$C_{\pi^{0}}=\frac{1}{2}$ and $C_{\eta}=\frac{2}{3}$ or
$\frac{1}{6}$. Different values of $C_{\eta}$ corresponds to the
initial state being $b\overline{u}, b\overline{d}$ or $
b\overline{s}$ respectively. All hadronic coupling constants depends
on the radial quantum number. For the decay within n=1 they are
notated as $g_{HH}$, $g_{SH}$ etc, and the decay from n=2 to n=1
they are represented by $\widetilde{g}_{HH}$, $\widetilde{g}_{SH}$,
 Higher order corrections for spin and
flavor violation of order $\frac{1}{m_{Q}}$ are excluded to avoid
new unknown coupling constants.

\section{Numerical Analysis}
The masses, decay widths and $J^{P'}s$ for the experimentally
available radial excited charm mesons $D_{0}(2560)$,
$D^{*}_{1}(2680)$, $D_{J}(3000)$, $D^{*}_{J}(3000)$ have been
analyzed with various theoretical approaches \cite{pallavi,25,26}.
In particular the predicted $J^{P}$ values for $(D_{J}(2560),
D^{*}_{J}(2680))$ are given as $(0^{-}, 1^{-})_{\frac{1}{2}}$ for
n=2 and L=0 and $J^{P}$ for states $(D^{*}_{J}(3000)),
(D_{J}(3000))$ are given as $(0^{+}, 1^{+})_{\frac{1}{2}}$ for n=2
and L=1.

 In bottom sector, only one radially excited bottom state i.e. $B_{J}(5970)$, is
 experimentally known, whose $J^{P}$ is associated with $1^{-}$
 \cite{13,28a}
for $\widetilde{B}^{*}_{1}(2S)$ state. Other radially excited bottom
states ($B(2S)$, $B(2P)$, $B_{s}(2S)$ and $B_{s}(2P)$) are still
unavailable. In this paper, we aim to predict the masses, decay
widths, branching ratios of these missing radially excited bottom
states B(2S) and B(2P) in the framework of HQET. To study the
behavior of the heavy-light mesons for their spectroscopy, masses
are the most important property to be studied so, we start our
calculations by predicting the masses of these bottom meson states.
To calculate these masses, we use the flavor symmetry property of
the heavy quarks $\lambda_{F}^{(b)} = \lambda_{F}^{(c)}$ and
$\Delta_{F}^{(b)} = \Delta_{F}^{(c)} $. This flavor symmetry
parameters are defined in terms of the spin averaged mass splittings
between the higher state doublets and the ground state doublet,
represented by $\Delta_{F}$ and $\lambda_{F}$ which is the mass
splittings between the spin partners of the doublets. From the LHCb
data \cite{lhcb2013}, spin averaged mass splittings $\Delta_{F}$ and
the hyperfine splittings $\lambda_{F}$ for the recently observed 2S
and 2P charm mesons states comes out to be:

\begin{center}
$\Delta_{\widetilde{H}}^{(c\overline{u})} = 660.33\pm3.8 MeV,\
\lambda_{\widetilde{H}}^{(c\overline{u})} = (213.43 \pm
3.9MeV)^{2}$\\
$\Delta_{\widetilde{S}}^{(c\overline{u})} = 1009.44\pm6.12 MeV,\
\lambda_{\widetilde{S}}^{(c\overline{u})} = (164.72
MeV\pm2.4)^{2}$\\
$\Delta_{\widetilde{T}}^{(c\overline{u})} = 1034.19\pm1.2 MeV,\
\lambda_{\widetilde{T}}^{(c\overline{u})} = (208.41
MeV\pm1.4)^{2}$\\
\end{center}
 for the n=2 odd parity, low lying even parity and
for the excited even parity $c\overline{u}$ mesons. The charm mesons
for n=2, P-wave with $j = 3/2$ are experimentally unavailable, so we
have taken the theoretical masses for $(\widetilde{D}_{1},
\widetilde{D}_{2}^{*})$ having values (2932.50, 3020.60)MeV
\cite{16, 27, 27a, 28a}. In this, we have taken the SU(2) isospin
symmetry for the non-strange charm mesons, i.e. M$(c\overline{u})$ =
M$(c\overline{d})$. The small statistical errors in $\Delta_{F}$ and
$\lambda_{F}$ ( $F = \widetilde{H}, \widetilde{S}, \widetilde{T}$)
for the non-strange radial excited charm mesons reflect the
precision of the LHCb results \cite{lhcb2013}.

Calculated masses obtained using these symmetries are listed in the
2nd column of Table \ref{nonbotmas}. Here, mass of bottom state
$\widetilde{B}_{1}^{*}$ for n=2 and $J^{P}$ = $1^{-}$  comes out to
be 5981.50 MeV, which is very close to experimentally observed mass
5978 MeV and 5969.20 MeV for bottom state $B_{J}(5970)$ observed by
CDF and LHCb collaboration respectively. Closeness in the
experimentally observed mass for bottom state
$\widetilde{B}_{1}^{*}$ and the mass obtained by HQET shows the
authenticity of this heavy quark symmetry. Since the experimental
information on radial excited bottom state is limited, the
authenticity of this symmetry cannot be completely justified just on
the basis on the one experimental available bottom state.

Based on these limitations, we have also compared the predicted
masses for other n=2 bottom states, listed in Table\ref{nonbotmas},
with some of the theoretically available data. The masses calculated
using the heavy quark symmetry in our work are in agreement with the
masses obtained by the potential model in Ref.\cite{28a}. Masses of
the non-strange bottom field $\widetilde{H}$ deviates by  0.4 \% and
0.5\%  when compared with the masses in Ref.\cite{28a} for
$\widetilde{B}_{0}$ and $\widetilde{B}_{1}^{*}$ states respectively.
Similar pattern is observed for P-wave masses where the deviations
are below 1\%. In contrary to this, these bottom masses are
deviating at most by 3.98\% with the theoretical data in
Ref.\cite{27a}

We have obtained a set of bottom spectra for n=2 using flavor
symmetry which upto a very good approximation matches with other
theoretical data as discussed above. \\Now, we would like to look
into the QCD and $1/m_{Q}$ corrections in the HQET lagrangian. QCD
and $1/m_{Q}$ corrections are applied to a scale of
$\Lambda_{QCD}/m_{Q}$, where they can be an important input to
decide the level of breaking of symmetry. The corrections to
$\Delta_{F}$ and $\lambda_{F}$ (where F = $\widetilde{H}$,
$\widetilde{S}$, $\widetilde{T}$) parameters are coming in the form
of
\begin{equation}
  \lambda_{F}^{(b)} =
\lambda_{F}^{(c)}\delta\lambda_{F} \qquad \Delta_{F}^{(b)} =
\Delta_{F}^{(c)}+\delta\Delta_{F}
\end{equation}
\\ The $\lambda_{F}$ parameter originates from the chromomagnetic
interaction, thus only QCD corrections are dominated and $1/m_{Q}$
effect is small. For applying the QCD corrections to the spin
hyperfine splitting relation $\lambda_{F}$, the values of the
parameter $\alpha_{s}(m_{b})$ and $\alpha_{s}(m_{c})$ are taken as
0.22 and 0.36 \cite{22}. The leading QCD correction to the
$\lambda_{F}$ relation comes out to be \\
\begin{center}$\delta(\lambda_{F}) =
(\frac{\alpha_{s}(m_{b})}{\alpha_{s}(m_{c})})^{9/25}$ = 0.83.
\end{center}
We then obtain:
\begin{multline}
\lambda_{\widetilde{H}}^{b} = (194.44\pm3.9)^{2} MeV^{2},\\
\lambda_{\widetilde{S}}^{b} = (150.06\pm2.19)^{2} MeV^{2},\\
\lambda_{\widetilde{T}}^{b} =(189.87\pm1.27)^{2} MeV^{2}
\end{multline}

\setlength{\tabcolsep}{0.09em} %
{\renewcommand{\arraystretch}{0.2}%
\begin{table*}[h]{\normalsize
\renewcommand{\arraystretch}{1.0}
\tabcolsep 0.2cm \caption{\label{nonbotmas}Predicted values of the
radially excited non-strange 2S and 2P bottom meson states. All the
masses are in MeV units}
 \noindent
\begin{tabular}{ccccccc}
\hline
 &$ 0^{-}(2^{1}S_{0})$ &$1^{-}(2^{3}S_{1}) $&$ 0^{+}(2^{3}P_{0})$ & $ 1^{+}(2^{1}P_{1})$&$ 1^{+}(2^{3}P_{1})$ &$
 2^{+}(2^{3}P_{2})$\\
 \hline
 \hline
 Without
 corrections&5950.96&5981.50&6335.83&6318.67&6336.30&6354.56\\
 Corrections
 in$\lambda_{F}$&5954.68&5980.26&6333.74&6319.37&6338.16&6353.45\\
Corrections
 in
 $\Delta_{F}$&5928$\pm13$&5959$\pm13$&6291$\pm37$&6274$\pm37$&6306$\pm9$&6324$\pm9$\\
 Correction in both parameters $\lambda_{F}$ and
 $\Delta_{F}$&5932$\pm13$&5957$\pm13$&6289$\pm37$&6274$\pm37$&6308$\pm9$&6323$\pm9$\\
Ref.\cite{27}&5985&6019&6264&6278&6296&6292\\
Ref.\cite{16}&5976&5992&6318&6321&6345&6359\\
Ref.\cite{27a}&5939&5966&6143&6153&6160&6170\\
Ref.\cite{28a}&6003&6029&6367&6375&6387&6382\\
\hline

\end{tabular}}
\end{table*}
  The calculated bottom meson masses inherited with this QCD correction
  are tabulated in the 3rd column of the Table \ref{nonbotmas}.
The results obtained with this correction are deviating by 1 or 2
MeV except for the $\widetilde{B}_{1}^{*}$ state, where the
deviation is of 4 MeV value. The experimental mass 5978 MeV observed
by CDF collaboration for bottom state $B_{J}(5970)$ is now deviating
by 2 MeV from the mass obtained by applying the QCD correction to
the $\lambda_{F}$. Thus the gap between experimental and theoretical
HQET mass has been reduced when the correction has been introduced.
While studying this correction, we realize that these kind of
correction for n=2 bottom mesons are showing similar pattern of
behavior but with a reduced effect, when compared with results for
n=1 bottom states. Specifically in Ref.\cite{27b}, most of the
bottom masses are deviating from their values by at 6-7 MeV. This is
something we expect when we study the properties of the heavier
mesons for higher excited states.

Now, in the next section, we will study the effect of QCD and
$1/m_{Q}$ corrections to other heavy flavor symmetry in the form of
$\delta\Delta_{F}$, where $F =
\widetilde{H},\widetilde{S},\widetilde{T}$. This kind of analysis is
already performed for n=1 states in Ref.\cite{21}, where the value
of $\delta\Delta_{\widetilde{S}} \sim \mathcal{O} (-35)$ MeV

 Since the experimental information on the radial
excited bottom state masses is unavailable, so we cannot predict the
exact value for these corrections $\delta\Delta_{F}$ for n=2.
However, by using the available theoretical masses for these radial
excited charm and bottom states, we have estimated a range to this
correction $\delta\Delta_{F}$ for n=2, that modifies $\Delta_{F}$
as:
\begin{multline}
  \Delta_{\widetilde{H}}^{b} = 638.58 \pm 13.75 MeV,\\
 \Delta_{\widetilde{S}}^{b} = 965.09 \pm 37.15 MeV,\\
  \Delta_{\widetilde{T}}^{b} = 1004.56 \pm 9.2 MeV,
\end{multline}
 Masses obtained by this correction are tabulated in 4th column of Table \ref{nonbotmas}. The
 Table shows that the correction $\delta\Delta_{F}$ to
 $\Delta_{F}^{(b)} = \Delta_{F}^{(c)}$ results in proportionate
  reduction in the bottom masses.

 Masses tabulated in the last column of the Table \ref{nonbotmas} are based on the corrections to both the parameters i.e.$\delta\lambda_{F}$ and $\delta\Delta_{F}$
 simultaneously. Since the effect of corrections to $\lambda_{F}$ is
 small, so the resultant masses are close to the masses obtained by
 applying corrections to the $\Delta_{F}$ parameter.

  In the strange sector, the experimentally known radially excited charm mesons are
   $D_{sJ}(2700)$ and $D_{sJ}(3040)$ \cite{5cs,6cs}.
   The $J^{P}$ for these states are theoretically \cite{30, 31, 32}
given as $(1^{-})_{\frac{1}{2}}$ (n=2,L=0) for $D_{sJ}(2700)$ and
$(1^{+})_{\frac{1}{2}}$ (n=2,L=1)for
  $D_{sJ}(3040)$ state.
The other strange 2S and 2P charm and bottom mesons are still
unknown. Here, we have predicted the unavailable 2S and 2P charm
masses using equation 24, and are listed in Table \ref{strangecharm}
and are matched with other theoretical data.

\setlength{\tabcolsep}{0.09em} %
{\renewcommand{\arraystretch}{0.2}%
\begin{table*}[h]{\normalsize
\renewcommand{\arraystretch}{1.0}
\tabcolsep 0.2cm \caption{\label{strangecharm} Predicted values of
the radially excited strange 2S and 2P charm meson states. All the
masses are in MeV units}
 \noindent
\begin{tabular}{|c|c|c|c|c|}

\hline

 $J^{P}(n^{2s+1}L_{J})$&Our&Ref.\cite{28a}&Ref.\cite{16}&Ref.\cite{28}\\

\hline
$ 0^{-}(2^{1}S_{0})$ &2682.96&2680&2688&2673\\

$1^{-}(2^{3}S_{1}) $&2754.63&2719&2731&2732\\

$ 0^{+}(2^{3}P_{0})$ &3007.80&3022&3054&3005 \\
 $ 1^{+}(2^{1}P_{1})$&3009.90&3081&3154&3018\\
$ 1^{+}(2^{3}P_{1})$ &3089.61&3092&3067&3038\\
 $ 2^{+}(2^{3}P_{2})$&3127.71&3109&3142&3048\\

 \hline
\end{tabular}}
\end{table*}

 In the same way, strange bottom masses are calculated and are
 reported
 in Table \ref{strangebottom}. The $2^{nd}$ column of the Table gives the masses without any
 correction and the masses listed in column $3^{rd}$ and $4^{th}$ include the
 corrections to the $\lambda_{F}$ and to $\Delta_{F}$ relation respectively.
  This is followed by the masses in
 $5^{th}$ column which gives the masses obtained by using
 corrections to both the $\lambda_{F}$ and $\Delta_{F}$ relations simultaneously.
  In general, QCD correction to $\lambda_{F}$ relation changes the bottom masses
 by few MeV. Correction to $\Delta_{F}$ relation results in deviating the mass
 of S-wave bottom states by 0.22\% . And the masses for P-wave get deviated
  by 0.77\% and 0.58\% respectively for $s_{l} = 1/2^{+}$ and $s_{l} =
 3/2^{+}$.

\setlength{\tabcolsep}{0.09em} %
{\renewcommand{\arraystretch}{0.2}%
\begin{table*}[h]{\normalsize
\renewcommand{\arraystretch}{1.0}
\tabcolsep 0.2cm \caption{\label{strangebottom}Predicted values of
the radially excited strange 2S and 2P bottom meson states. All the
masses are in MeV units}
 \noindent
 \begin{tabular}{ccccccc}
\hline
 &$ 0^{-}(2^{1}S_{0})$ &$1^{-}(2^{3}S_{1}) $&$ 0^{+}(2^{3}P_{0})$ & $ 1^{+}(2^{1}P_{1})$&$ 1^{+}(2^{3}P_{1})$ &$
 2^{+}(2^{3}P_{2})$\\
 \hline\hline
 Without
 corrections&6039.67&6071.85&6335.72&6336.72&6429.02&6447.41\\
 Corrections
 in$\lambda_{F}$&6043&6070.54&6335.84&6336.68&6430.89&6446.29\\
Corrections
 in
 $\Delta_{F}$&6025$\pm6$&6058$\pm6$&6286$\pm8$&6287$\pm8$&6391$\pm20$&6410$\pm20$\\
 Correction in both parameters $\lambda_{F}$ and
 $\Delta_{F}$&6029$\pm6$&6056$\pm6$&6286$\pm8$&6287$\pm8$&6393$\pm20$&6409$\pm20$\\
Ref.\cite{27}&5886&5920&6163&6175&6194&6188\\
Ref.\cite{16}&5890&5906&6221&6209&6281&6260\\
Ref.\cite{27a}&5822&5848&6010&6022&6028&6040\\
Ref.\cite{28a}&5926&5947&6297&6295&6311&6299\\
\hline
\end{tabular}}

\end{table*}

Now we study the various other properties of bottom mesons like
strong decay widths, branching ratios, strong coupling constants. We
apply the effective Lagrangian approach discussed in Sec II to
calculate the OZI allowed two body strong decay widths and the
various branching ratios involved with the  bottom states $B(2S)$,
$B(2P)$, $B_{s}(2S)$ and $B_{s}(2P)$. The numerical value of the
partial and total decay widths of 2S and 2P family are collected in
Table \ref{width1}, \ref{width2}, \ref{width3}, where Table
\ref{width1} and \ref{width2} are for the n=2, S and P wave bottom
meson with $s_{l} = 1/2$, and the Table \ref{width3} is for the
other P wave bottom meson states having $s_{l} = 3/2$.

\setlength{\tabcolsep}{0.09em} %
{\renewcommand{\arraystretch}{0.2}%

\begin{table*}[h]{\normalsize
\renewcommand{\arraystretch}{1.0}
\tabcolsep 0.2cm \caption{Strong decay width of non-strange and
strange n=2 S-wave bottom mesons $B(2 ^{3}S_{1})$, $B(2 ^{1}S_{0})$,
$B_{s}(2 ^{3}S_{1})$ and $B_{s}(2 ^{1}S_{0})$. Ratio in 5th column
represents the $\widehat{{\bf \Gamma}}=
\frac{\Gamma}{\Gamma(B_{J}^{(*)} \rightarrow B^{*+}\pi^{-})}$ for
the non-strange mesons and $\widehat{{\bf \Gamma}}=
\frac{\Gamma}{\Gamma(B_{sJ}^{*} \rightarrow B^{*0}K^{+})}$ for the
strange mesons. Fraction gives the percentage of the partial decay
width with respect to the total decay width. }\label{width1}
\begin{center}
\begin{tabular}{c|c|c|c|c|c}
\hline \hline State&$nLs_{l}J^{P}$&Decay channel&Decay
Width(MeV)&Ratio&Fraction\\
\hline
$\widetilde{B}_{1}^{*}(5981.50)$& 2$S_{1/2}1^{-}$&$B^{*}\pi^{+}$&1246.27$\widetilde{g}^{2}_{HH}$&1&37.97\\
&&$B^{*}\pi^{0}$&626.01$\widetilde{g}^{2}_{HH}$&0.50&19.07\\
&&$B^{*}\eta$&37.18$\widetilde{g}^{2}_{HH}$&0.02&1.13\\
&&$B^{*}_{s}K$&96.36$\widetilde{g}^{2}_{HH}$&0.07&2.93\\
&&$B^{0}\pi^{0}$&377.54$\widetilde{g}^{2}_{HH}$&0.30&11.50\\
&&$B^{+}\pi^{-}$&753.14$\widetilde{g}^{2}_{HH}$&0.60&22.94\\
&&$B^{0}\eta$&32.41$\widetilde{g}^{2}_{HH}$&0.02&0.98\\
&&$B_{s}K$&112.72$\widetilde{g}^{2}_{HH}$&0.09&3.43\\
&&Total&3281.67$\widetilde{g}^{2}_{HH}$&&\\
\hline

$\widetilde{B}_{0}(5950.96)$& 2$S_{1/2}0^{-}$&$B^{*}\pi^{+}$&1629.05$\widetilde{g}^{2}_{HH}$&1&64.16\\
&&$B^{*}\pi^{0}$&818.67$\widetilde{g}^{2}_{HH}$&0.50&32.24\\
&&$B^{*}\eta$&33.15$\widetilde{g}^{2}_{HH}$&0.02&1.30\\
&&$B^{*}_{s}K$&57.78$\widetilde{g}^{2}_{HH}$&0.03&2.27\\
&&Total&2538.67$\widetilde{g}^{2}_{HH}$&&\\
\hline
 $\widetilde{B}^{*}_{s}$(6071.85)&2$S_{s1/2}1^{-}$&$B^{0}K^{0}$&520.79$\widetilde{g}^{2}_{HH}$&0.65&13.24\\
&&$B^{+}K^{-}$&529.94$\widetilde{g}^{2}_{HH}$&0.66&13.47\\
&&$B_{s}\pi^{0}$&384.18$\widetilde{g}^{2}_{HH}$&0.48&9.77\\
&&$B_{s}\eta$&134.70$\widetilde{g}^{2}_{HH}$&0.16&3.42\\
&&$B^{*0}K^{0}$&784.86$\widetilde{g}^{2}_{HH}$&0.98&19.96\\
&&$B^{*+}K^{-}$&799.72$\widetilde{g}^{2}_{HH}$&1&20.34\\
&&$B^{*}_{s}\pi^{0}$&628.19$\widetilde{g}^{2}_{HH}$&0.78&15.97\\
&&$B^{*}_{s}\eta$&148.94$\widetilde{g}^{2}_{HH}$&0.18&3.78\\
&&Total&3931.35$\widetilde{g}^{2}_{HH}$&&\\
\hline
$\widetilde{B}_{s0}(6039.67)$&2$S_{s1/2}0^{-}$&$B^{*0}K^{0}$&924.75$\widetilde{g}^{2}_{HH}$&0.97&32.86\\
&&$B^{*+}K^{-}$&945.50$\widetilde{g}^{2}_{HH}$&1&33.60\\
&&$B^{*}_{s}\pi^{0}$&815.12$\widetilde{g}^{2}_{HH}$&0.86&28.97\\
&&$B^{*}_{s}\eta$&128.26$\widetilde{g}^{2}_{HH}$&0.13&4.55\\
&&Total&2813.64$\widetilde{g}^{2}_{HH}$&&\\
 \hline \hline
\end{tabular}
\end{center}}
\end{table*}
}
 For the radially ground state S-wave bottom states, Table \ref{width1} reveals $B^{*}\pi^{-}$ mode to be the dominant decay
 mode both for $\widetilde{B}_{1}^{*}$ and $\widetilde{B}_{0}$ bottom states with branching fraction of 37.97$\%$
 and 64.16$\%$ respectively. And for their strange partners, $B^{+}K^{-}$ is
 seen to be the leading decay mode with branching fraction 20.34$\%$
 and 33.60$\%$ for $\widetilde{B}_{s1}^{*}$ and $\widetilde{B}_{s0}$ state respectively. Hence the decay modes
 $B^{*}\pi^{-}$ and $B^{+}K^{-}$ are suitable for the experimental search for the missing
 non-strange and strange 2S bottom meson states. Total decay width for bottom state
 $\widetilde{B}_{1}^{*}$ is 64.32 MeV, which matches with its experimental value of 70 MeV
 \cite{11}(observed by CDF Collaboration), where strong coupling constant
  $\widetilde{g}_{HH}$ is used as 0.14 \cite{20}.
  This coupling  for the other S-wave bottom states gives decay width
 $\Gamma(\widetilde{B}_{0}) = 49.48$ MeV,
 $\Gamma(\widetilde{B}_{s1}^{*}) = 77.05$ MeV and
 $\Gamma(\widetilde{B}_{s0}) = 55.14$ MeV.
  \\Similarly, for n=2 low lying P-wave bottom states($0^{+},1^{+}$), Table \ref{width2}
   shows that the dominant decay modes for bottom state $\widetilde{B}_{0}^{*}$ and
  $\widetilde{B}_{1}^{'}$ are $B^{+}\pi^{-}$ and $B^{*+}\pi^{-}$ respectively.
   These decay modes contribute 42.15$\%$
 and 42.83$\%$ to the total
  decay widths of $\widetilde{B}_{0}^{*}$ and $\widetilde{B}_{1}^{'}$ state. And for the
  strange states $\widetilde{B}_{s0}^{*}$ and $\widetilde{B}_{s1}^{'}$, $B^{-}K^{+}$ and $B^{*}K^{+}$ decay
  modes emerges as the prominent modes for their experimental exploration in
  future. Using the coupling constant $\widetilde{g}_{SH}$ = 0.12,
  the total decay width for these $s_{l}$
  = 1/2 P-wave bottom states are obtained as:
$\Gamma(\widetilde{B}_{0}^{*}) =  242.70$ MeV,
 $\Gamma(\widetilde{B}_{1}^{'}) =  203.46$ MeV,
 $\Gamma(\widetilde{B}_{s0}^{*}) =  276.85$ MeV and
 $\Gamma(\widetilde{B}_{s1}^{'}) =  243.66$ MeV.

 Apart from the mentioned partial decay widths, these bottom states
 also decays to D-wave bottom mesons. But these decays are suppressed
  in our calculations because of their small contribution.

 \setlength{\tabcolsep}{0.09em} %
{\renewcommand{\arraystretch}{0.2}%

\begin{table*}{\normalsize
\renewcommand{\arraystretch}{1.0}
\tabcolsep 0.2cm \caption{Strong decay width of non-strange and
strange n=2 P-wave with $s_{l}$ = 1/2 bottom mesons $B(2
^{3}P_{0})$, $B(2 ^{1}P_{1})$, $B_{s}(2 ^{3}P_{0})$ and $B_{s}(2
^{1}P_{1})$. Ratio in 5th column represents the $\widehat{{\bf
\Gamma}}= \frac{\Gamma}{\Gamma(B_{J}^{(*)} \rightarrow
B^{*+}\pi^{-})}$ for the non-strange mesons and $\widehat{{\bf
\Gamma}}= \frac{\Gamma}{\Gamma(B_{sJ}^{*} \rightarrow B^{*0}K^{+})}$
for the strange mesons. Fraction gives the percentage of the partial
decay width with respect to the total decay width.}\label{width2}
\begin{center}
\begin{tabular}{c|c|c|c|c|c}
\hline \hline State&$nLs_{l}J^{P}$&Decay channel&Decay
Width(MeV)&Ratio&Fraction\\
\hline $\widetilde{B}_{0}^{*}(6335.83)$&2$P_{1/2}0^{+}$&$B^{-}\pi^{+}$&7087.60$\widetilde{g}^{2}_{SH}$&1&42.15\\
&&$B^{0}\pi^{0}$&3542.06$\widetilde{g}^{2}_{SH}$&0.49&21.06\\
&&$B^{0}\eta$&1063.91$\widetilde{g}^{2}_{TH}$&0.15&6.32\\
&&$B_{s}K$&5118.49$\widetilde{g}^{2}_{TH}$&0.72&30.44\\
&&Total&16812.10$\widetilde{g}^{2}_{SH}$&&\\
 \hline
$\widetilde{B}_{1}^{'}(6318.67)$&2$P_{1/2}1^{+}$&$B^{*}\pi^{+}$&6052.72$\widetilde{g}^{2}_{SH}$&1&42.83\\
&&$B^{*}\pi^{0}$&3027.69$\widetilde{g}^{2}_{SH}$&0.50&21.42\\
&&$B^{*}\eta$&890.83$\widetilde{g}^{2}_{TH}$&0.14&6.30\\
&&$B_{s}^{*}K$&4158.15$\widetilde{g}^{2}_{TH}$&0.68&29.42\\
&&Total&14129.40$\widetilde{g}^{2}_{SH}$&&\\
\hline
$\widetilde{B}_{0s}^{*}(6335.72)$&2$P_{s1/2}0^{+}$&$B_{s}\pi^{0}$&2839.71$\widetilde{g}^{2}_{SH}$&0.43&14.77\\
&&$B_{s}\eta$&3308.90$\widetilde{g}^{2}_{TH}$&0.50&17.21\\
&&$B^{+}K^{0}$&6530.27$\widetilde{g}^{2}_{SH}$&0.99&33.96\\
&&$B^{-}K^{+}$&6547.13$\widetilde{g}^{2}_{SH}$&1&34.05\\
&&Total&19226.00$\widetilde{g}^{2}_{SH}$&&\\
\hline
$\widetilde{B}_{1s}^{'}(6336.72)$&2$P_{s1/2}1^{+}$&$B^{*}_{s}\pi^{0}$&2490.50$\widetilde{g}^{2}_{SH}$&0.42&14.71\\
&&$B_{s}^{*}\eta$&2834.34$\widetilde{g}^{2}_{TH}$&0.48&16.75\\
&&$B^{*+}K^{0}$&$5792.74\widetilde{g}^{2}_{SH}$&0.99&34.23\\
&&$B^{*-}K^{+}$&$5803.61\widetilde{g}^{2}_{SH}$&1&34.29\\
&&Total&16921.20$\widetilde{g}^{2}_{SH}$&&\\
 \hline \hline
\end{tabular}
\end{center}}
\end{table*}
}
Lastly, for the other P-wave bottom states having $s_{l}$
  = 3/2, Table \ref{width3} points $B^{*}\pi^{+}$ and $B^{*}K^{-}$ decay modes to be the
  best suitable for the study of the missing non-strange states $(\widetilde{B}_{1},\widetilde{ B}_{2}^{*})$ and strange
  states $(\widetilde{B}_{s1}, \widetilde{B}_{s2}^{*})$
  respectively. While observing the total decay width values from
  Table, we notice that
  lower values of coupling constant $\widetilde{g}_{TH}$ will give realistic decay width value for these states.
  Assuming that the coupling constant $\widetilde{g}_{SH}$ and $\widetilde{g}_{TH}$
   will not vary much for higher excited states, total decay width corresponding to the coupling constant
   $\widetilde{g}_{TH}$ =0.12
  value are obtained as
$\Gamma(\widetilde{B}_{1}) = 188.69$ MeV, $
\Gamma(\widetilde{B}_{2}^{*}) = 226.39$ MeV,
 $\Gamma(\widetilde{B}_{s1}) = 259.89$ MeV and
 $\Gamma(\widetilde{B}_{s2}^{*}) = 313.98$ MeV.
 Thus the states $\widetilde{B}_{2}^{*}$ and $\widetilde{B}_{s2}^{*}$ are observe to be broader states as
 compared to their spin partner states $\widetilde{B}_{1}$ and $\widetilde{B}_{s1}$
 respectively. Here, we need to emphasize that the estimated total
 width of these states does not include the contribution from the
 decays to n=1 D-wave bottom mesons since the phase space is very
 small for these decay states.

\setlength{\tabcolsep}{0.09em} %
{\renewcommand{\arraystretch}{0.2}%

\begin{table*}{\normalsize
\renewcommand{\arraystretch}{1.0}
\tabcolsep 0.2cm \caption{Strong decay width of non-strange and
strange n=2 P-wave with $s_{l}$ = 3/2 bottom mesons $B(2
^{1}P_{1})$, $B(2 ^{3}P_{2})$, $B_{s}(2 ^{1}P_{1})$ and $B_{s}(2
^{3}P_{2})$. Ratio in 5th column represents the $\widehat{{\bf
\Gamma}}= \frac{\Gamma}{\Gamma(B_{J}^{(*)} \rightarrow
B^{*+}\pi^{-})}$ for the non-strange mesons and $\widehat{{\bf
\Gamma}}= \frac{\Gamma}{\Gamma(B_{sJ}^{*} \rightarrow B^{*0}K^{+})}$
for the strange mesons. Fraction gives the percentage of the partial
decay width with respect to the total decay width.}\label{width3}
\begin{center}
\begin{tabular}{c|c|c|c|c|c|c}
\hline \hline State&$nLs_{l}J^{P}$&Decay channel&Decay
Width(MeV)&Ratio&Fraction\\
\hline $\widetilde{B}_{1}(6336.30)$&2$P_{3/2}1^{+}$&$B^{*}\pi^{+}$&7021.89$\widetilde{g}^{2}_{TH}$&1&53.58\\
&&$B^{*}\pi^{0}$&3522.08$\widetilde{g}^{2}_{TH}$&0.50&26.87\\
&&$B^{*}\eta$&512.74$\widetilde{g}^{2}_{TH}$&0.07&3.91\\
&&$B_{s}K$&2046.74$\widetilde{g}^{2}_{HH}$&0.29&15.61\\
&&Total&13103.50$\widetilde{g}^{2}_{TH}$&&\\
 \hline
$\widetilde{B}_{2}^{*}(6354.56)$&2$P_{3/2}2^{+}$&$B^{*}\pi^{+}$&4571.90$\widetilde{g}^{2}_{TH}$&1&29.08\\
&&$B^{*}\pi^{0}$&2292.95$\widetilde{g}^{2}_{TH}$&0.50&14.58\\
&&$B^{*}\eta$&345.57$\widetilde{g}^{2}_{TH}$&0.07&2.19\\
&&$B^{*}_{s}K$&1392.05$\widetilde{g}^{2}_{TH}$&0.30&8.85\\
&&$B^{0}\pi^{0}$&1849.52$\widetilde{g}^{2}_{TH}$&0.40&11.76\\
&&$B^{+}\pi^{-}$&3694.17$\widetilde{g}^{2}_{TH}$&0.80&23.49\\
&&$B^{0}\eta$&300.86$\widetilde{g}^{2}_{TH}$&0.06&1.91\\
&&$B_{s}K$&1274.68$\widetilde{g}^{2}_{TH}$&0.27&8.10\\
&&Total&15721.70$\widetilde{g}^{2}_{TH}$&&\\
\hline
$\widetilde{B}_{1s}(6429.02)$&2$P_{s3/2}1^{+}$&$B^{*}_{s}\pi^{0}$&3587.85$\widetilde{g}^{2}_{TH}$&0.57&19.87\\
&&$B^{*+}K^{0}$&6149.63$\widetilde{g}^{2}_{TH}$&0.98&34.07\\
&&$B^{*-}K^{+}$&6212.01$\widetilde{g}^{2}_{TH}$&1&34.41\\
&&$B^{*}_{s}\eta$&2098.43$\widetilde{g}^{2}_{TH}$&0.33&11.62\\
&&Total&18047.90$\widetilde{g}^{2}_{TH}$&&\\
\hline
$\widetilde{}B_{2s}^{*}(6447.41)$&2$P_{s3/2}2^{+}$&$B^{*}_{s}\pi^{0}$&2336.94$\widetilde{g}^{2}_{TH}$&0.57&10.71\\
&&$B^{*+}K^{0}$&$4053.03\widetilde{g}^{2}_{TH}$&0.99&18.58\\
&&$B^{*-}K^{+}$&$4092.49\widetilde{g}^{2}_{TH}$&1&18.76\\
&&$B^{*}_{s}\eta$&1414.93$\widetilde{g}^{2}_{TH}$&0.34&6.48\\
&&$B^{+}K^{-}$&3387.56$\widetilde{g}^{2}_{TH}$&0.82&15.53\\
&&$B^{0}K^{0}$&3352.34$\widetilde{g}^{2}_{TH}$&0.81&15.37\\
&&$B_{s}\pi^{0}$&1912.00$\widetilde{g}^{2}_{TH}$&0.46&8.76\\
&&$B_{s}\eta$&1255.43$\widetilde{g}^{2}_{TH}$&0.30&18.58\\
&&Total&21804.70$\widetilde{g}^{2}_{TH}$&&\\
 \hline \hline
\end{tabular}
\end{center}}
\end{table*}
}
\section{Conclusion}

With so many newly available charm states, the data for the higher
excited bottom states is limited as compared to the charm sector. In
this work, we focuses on predicting the masses and the strong decay
widths of the experimentally missing radially excited bottom states
B(2S), B(2P), $B_{s}(2S)$ and $B_{s}(2P)$ using heavy quark
effective theory.\\
\begin{itemize}
  \item   We
apply the heavy quark symmetry property to the experimentally
available radially excited charm mesons observed by LHCb and
evaluate the similar spectra for the bottom sector. The predicted
mass of bottom state $\widetilde{B}_{1}^{*}$ in our work is only
0.33\% deviating from the experimental measured mass of
$B_{J}(5970)$ state. The Closeness in the experimentally observed
mass for bottom state $\widetilde{B}_{1}^{*}$ and the mass obtained
in our work shows the authenticity of this heavy quark symmetry.
\\ \item These masses
are then studied by taking the QCD and $1/m_{Q}$ corrections in the
form of $\delta\Delta_{F}$ and $\delta\lambda_{F}$. Masses obtained
using these corrections are tabulated in Table \ref{nonbotmas}.
While studying these correction, we realize that $\delta\lambda_{F}$
shifts the bottom masses at most by 0.06\% and has reduced the
existing gap between the experimental and the observed mass of
$\widetilde{B}_{1}^{*}$ state. While the correction
$\delta\Delta_{F}$ results in proportionate
  reduction in the bottom masses by the amount of $|\delta\Delta_{F}|$.
\\ \item Evaluating the strange bottom masses in the similar manner, we
discover that calculated masses are in good agreement with maximum
2.5\% deviation from the other theoretical stranged bottom masses.
Corrections results in deviating the S-wave masses by 0.22\%. And
the masses for P-wave get deviated
  by 0.77\% and 0.58\% respectively for $s_{l} = 1/2^{+}$ and $s_{l} =
 3/2^{+}$. The calculated strange bottom masses in this work for n=2 are
$\sim90$ MeV higher than the masses of the non-strange bottom masses
except for the low lying S-wave states. These kind of corrections
for n=2 bottom mesons are showing similar pattern of behavior but
with a reduced effect, when compared with results for n=1 bottom
states. This is something we expect when we study the properties of
the heavier mesons for higher excited states.
 \\
\item With these obtained masses, we further calculated the strong
decay widths of these bottom states, which are collected in Table
\ref{width1},\ref{width2},\ref{width3}. The predicted decay widths
are in the terns of strong coupling constants $\widetilde{g}_{HH}$
$\widetilde{g}_{SH}$ and $\widetilde{g}_{TH}$. To avoid these
unknown couplings, we have also computed the branching ratios and
fractions for the possible OZI allowed decay channels.
\\ \item While analysing the decay
widths in the Table \ref{width1},\ref{width2},\ref{width3}, we
observe $B^{*}\pi^{-}$ ($B^{+}K^{-}$) mode to be the dominant decay
 mode both for $\widetilde{B}_{1}^{*}$ ($\widetilde{B}_{s1}^{*}$)
 and $\widetilde{B}_{0}$ ($\widetilde{B}_{s0}$) bottom states.
Similarly, for the low lying P=wave bottom mesons
$\widetilde{B}_{0}^{*}$ and
  $\widetilde{B}_{1}^{'}$, $B^{+}\pi^{-}$ and $B*^{+}\pi^{-}$ emerges
   as the dominant decay modes with contribution of 42.15$\%$
 and 42.83$\%$ to their total
  decay widths. For their strange partners, $B^{-}K^{+}$ and $B^{*}K^{+}$ are seen as the
   prominent modes for
   $\widetilde{B}_{s0}^{*}$ and $\widetilde{B}_{s1}^{'}$
   states respectively. Lastly, for the excited P-wave bottom mesons,
    $B^{*}\pi^{+}$ and $B^{*}K^{-}$  are seen to be the
  best suitable decay modes for the study of the missing non-strange states
   $(\widetilde{B}_{1},\widetilde{ B}_{2}^{*})$ and strange
  states $(\widetilde{B}_{s1}, \widetilde{B}_{s2}^{*})$
  respectively. With the obtained decay widths, we have
 further check their sensitivity to the QCD and $1/m_{Q}$ corrected masses.
 It is found that these corrections shifts the decay width values at
 most by 30 MeV $(\sim12\%)$.
\end{itemize}
 Finally, we have predicted the masses and the strong decay widths
  of the experimentally not yet observed n=2 bottom mesons.
 These predicted bottom
states has opened a window to investigate the higher excitations of
bottom mesons and can be confronted with the future experimental
data at the LHCb, D0, CDF.

\end{document}